\documentstyle[12pt]{article}

\newenvironment{mylist}[1]{
\setbox1=\hbox{#1}
\begin{list}{}{
\setlength{\labelwidth}{\wd1}
\setlength{\leftmargin}{\wd1}
  \addtolength{\leftmargin}{1em}
  \addtolength{\leftmargin}{\labelsep}
\setlength{\rightmargin}{1em}}}{\end{list}}

\newcommand{\litem}[1]{\item[#1\hfill]}

\makeatletter
\def\maketitle{\par
 \begingroup
 \def\thefootnote{\arabic{footnote}}    
 \def\@makefnmark{\hbox to 0pt{$^{\@thefnmark}$\hss}} \if@twocolumn \twocolumn[\@maketitle] \else \newpage
 \global\@topnum\z@ \@maketitle \fi\thispagestyle{plain}\@thanks
 \endgroup
 \let\maketitle\relax
 \let\@maketitle\relax
 \gdef\@thanks{}\gdef\@author{}\gdef\@title{}\let\thanks\relax}
\makeatother
\begin{document}

\sloppy

\newcommand{\ignore}[1]{}
\newcommand{\remark}[1]{[{\small{\bf Editorial Remark: }#1}]}

\newtheorem{theorem}{Theorem}
\newtheorem{lemma}[theorem]{Lemma}
\newtheorem{conjecture}[theorem]{Conjecture}
\newtheorem{corollary}[theorem]{Corollary}
\newtheorem{definition}[theorem]{Definition}
\newcommand{\abs}[1]{| #1 |}

\def\sl{\it}
\def\acsc{\it}
\def\focs#1{Proceedings of the #1 Annual Symposium on the Foundations of
Computer Science}
\def\stoc#1{Proceedings of the #1 Annual ACM Symposium on Theory of Computing}

\def\qed{\vspace{.3em}\noindent\fbox{\rule{
0em}{.3em}\rule{.3em}{0em}}\vspace{1em}}

\def\proof{\noindent{\it Proof.}\ }
\def\sketch{\noindent{\it Proof sketch.}\ }

\def\th{^{th}}
\def\st{^{st}}

\begin{titlepage}

\title{\Large Competitive Paging Algorithms}
\author{{\it Amos Fiat} \thanks{
Department of Computer Science, Tel Aviv University, Tel Aviv, Israel 69978.
Support was provided by a Weizmann fellowship.
} \\
                     \and
{\it Richard M. Karp} \thanks{
Department of Electrical Engineering and Computer Science, Computer Science
Division, University of California, Berkeley, CA 94720.  Partial support was
provided by the International Computer Science Institute, Berkeley, CA,
and by NSF grant CCR-8411954.} \\
                     \and
{\it Michael Luby} \thanks{
Department of Computer Science, University of Toronto, Toronto, Ontario,
Canada M6C 3B7.  Support was provided by the International Computer Science
Institute and operating grant A8092 of the
Natural Sciences and Engineering Research Council of
Canada.  Current address: International Computer
Science Institute, Berkeley, CA 94704.} \\
                     \and
{\it Lyle A. McGeoch} \thanks{Department of Mathematics and Computer Science,Amherst College,
Amherst, MA 01002.} \\
                     \and
{\it Daniel D. Sleator} \thanks{
School of Computer Science, Carnegie Mellon University, Pittsburgh, PA
15213.  Partial support was provided by DARPA, ARPA order 4976, Amendment
20, monitored by the Air Force Avionics Laboratory under contract
F33615-87-C-1499, and by the National Science Foundation under grant
CCR-8658139.} \\
                     \and
{\it Neal E. Young} \thanks{
Computer Science Department, Princeton University, Princeton NJ 08544.  Part
of this work was performed while the author was at the Digital Equipment
Corp. Systems Research Center, Palo Alto, CA.} \\
}
\date{{\small \today}}
\end{titlepage}
\maketitle


\begin{abstract}

The {\it paging problem\/} is that of deciding which pages to keep in
a memory of $k$ pages in order to minimize the number of page faults.
We develop the {\it marking algorithm\/}, a randomized on-line
algorithm for the paging problem.  We prove that its expected cost on
any sequence of requests is within a factor of $2H_k$ of optimum.
(Where $H_k$ is the $k$th harmonic number, which is roughly $\ln k$.)
The best such factor that can be achieved is $H_k$.  This is in
contrast to deterministic algorithms, which cannot be guaranteed to be
within a factor smaller than $k$ of optimum.

An alternative to comparing an on-line algorithm with the optimum off-line
algorithm is the idea of comparing it to several other on-line algorithms.
We have obtained results along these lines for the paging problem.  Given a
set of on-line algorithms and a set of appropriate constants, we describe a
way of constructing another on-line algorithm whose performance is within
the appropriate constant factor of each algorithm in the set.

\vspace*{2em}

\end{abstract}

\section{Introduction}

Consider a memory system with $k$ pages of fast memory (a cache) and
$n-k$ pages of slow memory.  A sequence of requests to pages is to be
satisfied, and in order to satisfy a request to a page that page must be
in fast memory.  If a requested page is not in fast memory a {\it
page fault\/} occurs.  In this case the requested page must be moved
into fast memory, and (usually) a page must be moved from fast memory
to slow memory to make room for the requested page.  The {\it paging
problem\/} is that of deciding which page to eject from fast memory.
The cost to be minimized is the number of page faults.

A paging algorithm is said to be {\it on-line\/} if its decision of
which page to eject from fast memory is made without knowledge of
future requests.  Sleator and Tarjan \cite{st} analyzed on-line paging
algorithms by comparing their performance on any sequence of requests
to that of the optimum {\it off-line\/} algorithm (that is, one that
has knowledge of the entire sequence of requests in advance).  They
showed that two strategies for paging (ejecting the least recently
used page, or LRU, and first-in-first-out, or FIFO) could be worse
than the optimum off-line algorithm by a factor of $k$, but not more,
and that no on-line algorithm could achieve a factor less than $k$.

Karlin {\em et al.} \cite{kmrs} introduced the term {\it
competitive\/} to describe an on-line algorithm whose cost is
within a constant factor (independent of the request sequence) of the
optimum off-line algorithm, and they used the term {\it strongly competitive\/}
to describe an algorithm whose cost is within the
smallest possible constant factor of optimum.  These authors proposed
another paging strategy, {\it flush-when-full\/} or FWF, and showed
that it is also strongly $k$-competitive.

Manasse {\em et al.} \cite{mms} extended the definition of
competitiveness to include randomized on-line algorithms (on-line
algorithms which are allowed to use randomness in deciding what to
do).  Let $A$ be a randomized on-line algorithm, let $\sigma$ be a
sequence of requests, and let $\overline{C_A}(\sigma)$ be the
cost of algorithm $A$ on sequence $\sigma$ averaged over all
the random choices that $A$ makes while processing $\sigma$.  Let
$C_B(\sigma)$ be the cost of deterministic algorithm $B$ on sequence
$\sigma$.  Algorithm $A$ is said to be {\it $c$-competitive\/} if
there is a constant $a$ such that for every request sequence $\sigma$
and every algorithm $B$:
\[
\overline{C_A}(\sigma) \; \le \; c \cdot C_B(\sigma) + a.
\]
The constant $c$ is known as the {\it competitive factor\/}.
This definition has the desirable feature of ensuring that $A$'s
average performance on every individual sequence
is close to that of the optimum off-line algorithm.

In this paper we consider randomized algorithms for the paging problem from
the competitive point of view.  We describe a randomized algorithm, called
the {\it marking algorithm\/}, and show that it is $2H_k$-competitive.
(Here $H_k$ denotes that $k\th$ harmonic number:  $H_k = 1+{1\over 2} +
{1\over 3} + \cdots + {1\over k}$.  This function is closely approximated by
the natural logarithms: $\ln(k+1) \leq H_k \leq \ln(k)+1$.  We
also show that no randomized paging
algorithm can have a competitive factor less than $H_k$.

The marking algorithm is strongly competitive (its competitive factor
is $H_k$) if $k=n-1$, but it is not strongly competitive if $k<n-1$.  We
describe another algorithm, EATR, which is strongly competitive for
the case $k=2$.
Borodin, Linial, and Saks \cite{blspaper} gave the first specific
problem in which the competitive factor is reduced if the on-line
algorithm is allowed to use randomness.  The problem they analyzed is
the {\it uniform task system\/}.  They presented a randomized algorithm for
uniform task systems whose competitive factor is $2H_n$, where $n$ is
the number of states in the task system, and proved that for this
problem the competitive factor of any randomized algorithm is at least
$H_n$.  The marking algorithm is an adaptation of the randomized
algorithm of Borodin, {\it et al.\/} It was discovered by three groups
working independently.  These three groups collaborated in the writing
of this paper.

\vspace{1em}

The standard definition of competitiveness requires that the on-line
algorithm be within a constant factor of any other algorithm, even an
off-line one.  In the case of deterministic paging algorithms this
constraint is so severe that the best possible constant required is
rather large.  An alternative approach is to require that the on-line
algorithm be efficient compared to several other on-line algorithms
simultaneously.  Given deterministic on-line algorithms for the paging
problem $B(1), B(2), \ldots, B(m)$, and constants $c(1), c(2),\ldots,
c(m)$, we show how to construct a new on-line algorithm whose
performance is within a factor of $c(i)$ of $B(i)$ for all $1\leq i
\leq m$, under the condition that $1/c(1) + \cdots + 1/c(m) \leq 1$.  For
example, we can construct an algorithm whose performance is within a factor
of two of the performance of both the LRU algorithm and the FIFO algorithm.
We also show how this construction can be applied to randomized algorithms.

\vspace{1em}

This paper is organized as follows.  Section 2 defines {\it server
problems\/} (a generalized form of the paging problem) and introduces
the terminology we shall use for the paging problem.  Section 3 discusses
the marking algorithm, Section 4 describes algorithm EATR,
Section 5 proves the $H_k$ lower bound on the competitive factor, and
Section 6 contains our results about combining algorithms.  Recent extensions
to this work are described in Section 7, along with several open problems.

\section{Server problems}

To put our work on paging in context it is useful to point out the
connection between the paging problem and the {\it $k$-server
problem\/}.  Let $G$ be an $n$-vertex graph with positive edge lengths
obeying the triangle inequality, and let $k$ mobile servers occupy
vertices of $G$.  Given a sequence of requests, each of which
specifies a vertex that requires service, the $k$-server problem is to
decide how to move the servers in response to each request.  If a
requested vertex is unoccupied, then some server must be moved there.
The requests must be satisfied in the order of their occurrence in the
request sequence.  The cost of handling a sequence of requests is
equal to the total distance moved by the servers.

Server problems were introduced by Manasse, McGeoch and Sleator
\cite{mms,mmspaper}.  They showed that no deterministic algorithm for the
$k$-server problem can be better than $k$-competitive, they gave
$k$-competitive algorithms for the case when $k=2$ and $k=n-1$, and
they conjectured that there exists a $k$-competitive $k$-server
algorithm for any graph.  This conjecture holds when the graph is
uniform~\cite{st}, a weighted cache system (where the cost of moving
to a vertex from anywhere is the same)~\cite{ckpv}, a
line~\cite{ckpv}, or a tree~\cite{cl}.  Fiat {\em et al.} \cite{frr}
showed that there is an algorithm for the $k$-server problem with a
competitive factor that depends only on $k$.  There has also been
work on {\it memoryless\/} randomized algorithms for server
problems~\cite{bkt,cdrs,rs}.  These algorithms keep no information
between requests except the server locations.  The randomized
algorithm of Coppersmith {\em et al.}~\cite{cdrs} is $k$-competitive
for a large class of graphs.

In the {\it uniform $k$-server problem\/} the cost of moving a server
from any vertex to any other is one.  The paging problem is isomorphic
to the uniform $k$-server problem.  The correspondence between the two
problems is as follows: the pages of address space correspond to the
$n$ vertices of the graph, and the pages in fast memory correspond to
those vertices occupied by servers.  In the remainder of this paper we
shall use the terminology of the uniform $k$-server problem.

\section{The marking algorithm}

The marking algorithm is a randomized algorithm for the uniform
$k$-server problem on a graph with $n$ vertices.  The algorithm works
as follows.  The servers are initially on vertices $1, 2, 3, \ldots
k$.  The algorithm maintains a set of {\it marked\/} vertices.
Initially the marked vertices are exactly those that are covered by
servers.  After each request, the marks are updated, then a server is
moved if necessary, as follows:
\begin{mylist}{Marking: }
\litem{Marking: } Each time a vertex is requested, that vertex is marked.
The moment $k+1$ vertices are marked, all the marks except the one on the
most recently requested vertex are erased.
\litem{Serving: } If the requested vertex is already covered by a server,
then no servers move.  If the requested vertex is not covered, then a server
is chosen uniformly at random from among the unmarked vertices, and this
server is moved to cover the requested vertex.
\end{mylist}

This algorithm can be interpreted as a randomized form of LRU as
follows.  Rather than maintaining one queue of servers, the algorithm
maintains two of them.  When a server is needed it is taken from the
front of one of the queues, and placed at the end of the other.  When
the queue from which servers are taken is empty it is replaced by the
other queue, but not before the order of the elements in the queue is
shuffled by a random permutation.

\begin{theorem}
\label{markingthm1}
The marking algorithm is a $2H_k$-competitive algorithm for the uniform
$k$-server problem on $n$ vertices.
\end{theorem}

\proof
Let $\sigma = \sigma(1), \sigma(2), \ldots$ be a sequence of requests.  The
marking algorithm (denoted $M$) implicitly divides $\sigma$ (excluding
some requests at the beginning) into {\it phases\/}.  The first phase begins
with $\sigma(i)$, where $i$ is the smallest integer such that $\sigma(i)
\not\in \{1,2,\ldots,k\}$.  In general the phase starting with $\sigma(i)$
ends with $\sigma(j)$, where $j$ is the smallest integer such that the set
$\{\sigma(i),\sigma(i+1), \ldots, \sigma(j+1)\}$ is of cardinality $k+1$.

At the start of every phase, the marked vertices are precisely the
ones occupied by $M$'s servers.  The first request of every phase is
to an unmarked vertex. A vertex is called {\it clean\/} if it was not
requested in the previous phase, and has not yet been requested in
this phase.  A vertex is called {\it stale\/} if it was requested in
the previous phase, but has not yet been requested in this phase.

\ignore{The {\it clean\/} vertices of a phase are those that
are requested during the phase, but were not requested in the previous
phase.  Alternatively, the clean vertices are the vertices occupied by
servers at the end of the phase but not at the beginning of the phase.
A first request in the phase to a vertex that is not clean is said to be
{\it stale\/}. }

Our proof is organized as follows.  We let an adversary choose any
algorithm $A$.  We then evaluate the cost incurred by $A$ during a
phase, evaluate the cost incurred by $M$ during the same phase, and
compare these two quantities.  These costs depend on $l$, the number
of requests to clean vertices during the phase.

Without loss of generality,we shall assume that the algorithm $A$ chosen by the
adversary is {\it lazy\/}.  A lazy algorithm is one which does not
move any server in response to a request to a covered vertex and moves
exactly one server in response to a request to an uncovered vertex.
Manasse {\em et al.} \cite{mms,mmspaper} showed that for any given
algorithm, there is always a lazy one that incurs no more cost.  Thus
our assumption does not limit the generality of our result.

We shall first argue that the amortized cost incurred by $A$ over the
phase is at least $l/2$.  Let $d$ be the number of $A$'s servers that
do not coincide with any of $M$'s servers at the beginning of the
phase.  Let $d'$ be this quantity at the end of the phase.  Let $C_A$
be the cost incurred by $A$ in the phase.  We claim $C_A \geq l-d$,
because among the $l$ requests to clean vertices at most $d$ of
these will be for vertices that $A$ already covers.

A second bound on $C_A$ is obtained by considering $S$, the set of
marked vertices at the end of the phase.  The vertices of $S$ are
those that are covered by $M$ at the end of the phase, so at the end of
the phase $d'$ servers of $A$ are not in $S$.  During this phase
exactly the vertices of $S$ were requested, so since $A$ is lazy, we
know that at least $d'$ of $A$'s servers were outside of $S$ during
the entire phase.  The remaining $k-d'$ servers had to cover requests
at each of the $k$ vertices of $S$, implying that $A$'s cost is at
least $d'$.  That is, $C_A \geq d'$.

Combining the inequalities from the preceding paragraphs we get
$$
C_A ~ \geq ~ \max(l-d, d') ~ \geq ~ {1\over 2}(l-d+d').
$$
When this is summed over all phases, the $d$ and $d'$ terms telescope, so we
can assume for the purposes of this analysis that the cost of a phase is
just $l/2$.

We shall now bound the expected cost incurred by $M$ during the phase.
There are $l$ requests to clean vertices and each of these costs one.
There are $k-l$ requests to stale vertices, the expected cost of each
of these requests is just the probability that there is no server
there.  This probability varies as a function of the current number of
stale vertices, $s$, and the number of clean vertices requested in the
phase so far, $c$.  The expected cost of the request is $c/s$ because
there are $c$ unserved vertices distributed uniformly among $s$ stale
vertices.

During the phase, the sequence will make $l$ requests to clean
vertices and $k-l$ requests to stale vertices.  The sequence with the
highest expected cost for $M$ is the one which first requests all the clean
vertices (increasing $c$), before requesting any stale vertices.
The expected cost of the requests to stale vertices is thus bounded by
$$
{l\over k}+{l\over {k-1}}+{l\over {k-2}} \cdots + {l\over{l+1}}
= l(H_k - H_l).
$$
The total expected cost to $M$ for the phase is therefore at most
$$
l(H_k - H_l+1) \leq l H_k .
$$
Since the cost incurred by $A$ during the phase is (amortized)
$l/2$, this proves that the marking algorithm is $2H_k$-competitive.

\qed

In the special case of $n-1$ servers, we can obtain a tighter bound.

\begin{theorem}
\label{markingthm2}
The marking algorithm is a $H_{n-1}$-competitive algorithm for the uniform
$(n-1)$-server problem on $n$ vertices.
\end{theorem}

\proof The above proof can be modified slightly to give this theorem.  In
the $(n-1)$-server problem, every phase has $l=1$.  In this case we can show
that the amortized cost of $A$ is at least $l$ per phase.

As above, let $d$ be the number of $A$'s servers that do not coincide with
any of $M$'s servers at the beginning of the phase, and $d'$ be this
quantity at the end of the phase.  The first request of the phase is to a
clean vertex, and its cost to $A$ is at least $1-d$. Among the $n-2$ other vertices
requested in this phase, at least $d'$ cause $A$ to incur a cost of one.
Thus the cost to $A$ is at least $1-d+d'$.  This shows that the amortized
cost to $A$ of a phase is at least 1.  Combining this with the preceding
analysis of $M$ finishes the proof.

\qed

Why is it that if $l=1$ we can show that the cost to $A$ is at least
$l-d+d'$, but when $l>1$ we can only show that the cost is at least$(l-d+d')/2$?  The distinction is due to a difference in the structure of the
requests in a phase.  The tighter bound actually holds whenever the phase
has the following structure:  after the last request to a clean vertex, all
of the other $k-1$ vertices used during the phase are requested.  In this
case $A$ incurs a cost of $l-d$ for the clean vertices, then an additional
cost of $d'$ for the subsequent requests to the other $k-1$ vertices.  This
pattern holds for the case $l=1$.

The marking algorithm is not in general $H_k$-competitive for the
uniform $k$-server problem.  This is even true in the case $k=2$,
$n=4$.  Suppose that the servers of $M$ are initially on vertices 1
and 2, and the servers of the adversary $A$ are on vertices 1 and 3.
The first phase will consist of requests to vertices 3 and 4.  The
marking algorithm will incur a cost of 2 for these requests, and end
with servers on vertices 3 and 4.  To handle this phase, algorithm $A$
will use its server on vertex 3 to cover the request on vertex 4,
incurring a cost of 1.  At the end of the phase the servers of $A$ and
$M$ will again coincide on exactly one vertex, and the process can be
repeated.  The competitive factor for this application of $M$ is 2,
which exceeds $H_2$.

\section{Algorithm EATR}

Algorithm EATR is a randomized algorithm for the uniform $2$-server problem.
(The name stands for ``end after twice requested,'' a rough description of
how the algorithm defines the end of a phase.)

The servers are initially located on vertices $1$ and $2$.  The
algorithm partitions the sequence of requests into phases in a way
that is different from that used by the marking algorithm.  The first
phase starts at the first request that is to neither $1$ nor $2$.  A
vertex is called {\it clean\/} if it was not occupied by a server at
the end of the previous phase, and has not been requested during this
phase.  A vertex is called {\it stale\/} if it is not clean and is not
the most recently requested vertex.  The algorithm maintains one
server on the most recently requested vertex, and the other uniformly
at random among the set of stale vertices.  When a stale vertex is
requested the servers are placed on the two most recently requested
vertices.  The next phase begins after this, on a request for a vertex
that is not covered by a server.

\begin{theorem}
Algorithm EATR is a $3/2$-competitive algorithm for the uniform
$2$-server problem.
\end{theorem}

\proof
Let $l$ be the number of clean vertices requested during a phase.  Before the
request to the stale vertex that terminates the phase, the number of stale
vertices is $l+1$, and there is a server on each of these with probability
$1/(l+1)$.  The expected cost of a phase to EATR is then
$$
l + {l\over l+1} .
$$

The phases as defined by EATR have the special structure described in
the paragraph after the proof of Theorem \ref{markingthm2}.  Thus the
amortized cost incurred by any algorithm for a phase is at least
$l$.  The competitive factor is therefore at most
$$
{l+{l\over l+1} \over l} = 1+{1\over (l+1)} \leq {3\over 2}.
$$

\qed

\section{A lower bound}

\begin{theorem}
There is no $c$-competitive randomized algorithm forthe uniform $(n-1)$-server problem on $n$ vertices with
$c < H_{n-1}$.
\end{theorem}

\proof\footnote{Raghavan (\cite{raghavan}, pages 118--9)
presents a different proof of this theorem based on a
generalization of the minimax principle due to Andy Yao
\cite{yao}.}
Let $A$ be a randomized on-line algorithm for solving the problem.  We
use the technique of constructing a nemesis sequence for algorithm
$A$.  Since $A$ is randomized, the adversary constructing the sequence
is not allowed to see where the servers are.  The adversary is however
able to maintain a vector $p=(p_1, p_2, \ldots, p_n)$ of
probabilities, where $p_i$ is the probability that vertex $i$ is {\it
not\/} covered by a server.  (The adversary can do this by simulating
$A$ on on all possible outcomes of its random choices, and condensing
the information about where the servers are in each of these
simulations into the vector of probabilities.)  Note that $\sum_i p_i
= 1$.

If the nemesis sequence requests a vertex $i$, then the expected cost
incurred by $A$ is $p_i$.  As a result of responding to the request, $p_i$
changes to 1, and some other elements of the probability vector may decrease.
(Even if we allow $A$ to change the vector arbitrarily and ignore the cost
it incurs in doing so, the lower bound still holds.)

The adversary will maintain a set of {\it marked} vertices for the sequence
it has generated so far in just the way that the marking algorithm would.
Furthermore, we can also define phases in the nemesis sequence just as we
did for an arbitrary sequence processed by the marking algorithm.  As usual,
at the start of each phase $n-1$ vertices are marked.  After the first
request of the phase, one vertex is marked.

Armed with these tools (the marking and the probability vector), the
adversary can generate a sequence such that the expected cost of each phase
to $A$ is $H_{n-1}$, and the cost to the optimum off-line algorithm is 1.
This will prove the theorem.

Consider a situation in which the number of unmarked vertices is $u$.  The
goal of the adversary is to generate some requests that cause $A$ to incur
an expected cost of at least $1/u$ and decrease the number of unmarked
vertices to $u-1$ (except if $u=1$, in which case the number of unmarked
vertices changes to $n-1$).  Since $u$ takes on every integer value between
$1$ and $n-1$ the total expected cost incurred by $A$ is at least $H_{n-1}$.
This subsequence of requests will be called a {\it subphase\/}.
Constructing a subphase will show how to generate the desired nemesis
sequence, and complete the proof of the theorem.

A subphase consists of zero or more requests to marked vertices, followed by
a request to an unmarked vertex.  Let $S$ be the set of marked vertices, and
let $P = \sum_{i\in S} p_i$.  Let $u$ be $n-|S|$, the number of unmarked
vertices.  If $P=0$ then there must be an unmarked vertex $i$ with
$p_i \geq 1/u$.  In this case the subphase consists of a single request to
$i$.  The expected cost of this request is at least $1/u$.

If $P > 0$, then there must be $i\in S$ such that $p_i > 0$.  Let
$\epsilon=p_i$, and let the first request of the subphase be $i$.
Next, a set of requests are generated by the the following
loop: ($P$ denotes the current total probability of the marked vertices.)
\begin{quote}
While $P > \epsilon$, and while the total expected
cost of all the requests in this
subphase so far does not exceed $1/u$, request vertex $i\in S$, where
$p_i = \max_{j\in S}(p_j)$.
\end{quote}
Each iteration of this loop adds at least $\epsilon/|S| > 0$ to the total
expected cost of this subphase.  Thus the loop must terminate.
If the total expected cost ends up exceeding $1/u$, then an arbitrary
request is made to an unmarked vertex, and the subphase is over.  If the loop
terminates with $P \leq \epsilon$, then a request is generated to the
unmarked vertex $j$ with the highest probability value.
Note that $p_j \geq (1-P)/u$.  The following inequalities finish the proof:
$$
\mbox{expected cost of the subphase} \geq\epsilon + p_j\; \geq\; \epsilon + {1-P\over u} \geq\epsilon + {{1-\epsilon}\over u} \; \geq \; {1\over u}.
$$

\qed

If there are $k$ servers, with $1 \leq k \leq n-1$, then
the adversary can ignore all but $k+1$ vertices of the graph,
and force the on-line algorithm to incur a cost at least $H_k$ times
optimum.  Thus we have:

\begin{corollary}
There is no $c$-competitive randomized algorithm forthe uniform $k$-server problem on a graph of $n$ vertices with $c < H_k$,
where $1 \leq k \leq n-1$.
\end{corollary}

\section{Algorithms that are competitive against several others}

In many applications of the $k$-server model, the following situation
arises: one is given several on-line algorithms with desirable
characteristics, and would like to construct a single on-line
algorithm that has the advantages of all the given ones.  For example,
in the case of the paging problem (the uniform-cost $k$-server
problem) the least-recently-used page replacement algorithm (LRU) is
believed to work well in practice, but, in the worst case, can be $k$
times as costly as the optimal off-line algorithm; on the other hand,
we have exhibited a randomized on-line algorithm that is
$2H_k$-competitive, and thus has theoretical advantages over LRU. Can
we construct an on-line algorithm that combines the advantages of
these two algorithms? We shall see that the answer is ``Yes.''

We adopt the viewpoint that each on-line algorithm is tailored for a
particular choice of $k$, the number of servers, and $n$, the number
of vertices that may request service.  We may assume without loss of
generality that the $n$ vertices are named by the integers
$1,2,\ldots,n$.  The ordered pair $(k,n)$ is called the {\it type\/}
of the algorithm.  Thus, in a request sequence presented to an
algorithm of type $(k,n)$, each request is an integer between $1$ and
$n$.  According to this viewpoint a general strategy (such as LRU,
FIFO or the marking algorithm) determines infinitely many individual
algorithms, corresponding to all the possible choices of $k$ and $n$.

Let $A$ and $B$ be deterministic on-line algorithms of the same type.  Let
$c$ be a positive constant.  Then $A$ is said to be {\it $c$-competitive\/}
against $B$ if there exists a constant $a$ such that on every sequence
$\sigma$ of requests,$$
C_A(\sigma) \leq c\cdot C_B(\sigma) + a.
$$
Let $c^* = (c(1),c(2),\ldots,c(m))$ be a sequence of positivereal numbers.  Then $c^*$ is said to be {\it realizable\/} if, for everytype $(k,n)$, and for every sequence $B(1),B(2),\ldots,B(m)$ ofdeterministic on-line algorithms of type $(k,n)$, there exists adeterministic on-line algorithm $A$ of type $(k,n)$ such that, for
$i = 1,2,\ldots,m$, $A$ is $c(i)$-competitive against $B(i)$.

\begin{theorem}
\label{realizabletheorem1}
The sequence $c^*$ is realizable if and only if
\begin{equation}
\label{X}
\sum_{1\leq i \leq m}{1\over c(i)} \leq 1.
\end{equation}
\end{theorem}

\proof (Sufficiency) We show that, if (\ref{X}) holds, then $c^*$ is
realizable. Let deterministic on-line algorithms $B(1),B(2),\ldots,B(m)$ of
type $(k,n)$ be given. We shall construct a deterministic on-line algorithm
$A$ of type $(k,n)$ such that, for all positive integers $r$, all request
sequences $\sigma$, and all $i$ between 1 and $m$, $B(i)$ incurs a cost
greater than or equal to $\lfloor r/c(i) \rfloor$ by the time $A$ incurs
cost $r$. This will prove the sufficiency of (\ref{X}).

\ignore{This is an obsolete note to
note is for Fiat, Karp, and Luby.  There is a problem here.  In
order for the ``necessity'' part of the proof below to go through (the part
that states that if $1/c(1) + \cdots + 1/c(m) > 1$ then the sequence is not
realizable), this definition must be clarified.  The definition is not
analogous to the standard definition of competitiveness, because the
algorithms $A$ and $B$ must be defined for all problem sizes (values of $k$
and $n$), and the inequality must hold for all problem sizes, and all
sequences of requests.  Clearly you get into big trouble if you don't
require this.  For example, suppose you're considering the 2 server problem.
The vector (2,2,2) is then realizable, because LRU is 2-competitive, so it's
within a factor of 2 of any three other algorithms.  But $1/2+1/2+1/2 > 1$.

Rather than fixing the definition of ``$A$ being $c$-competitive against
$B$'', and thus making it non-analogous to the standard definition of
competitiveness, I propose that we simply drop the ``necessity'' part of the
theorem all together.

Another thing in favor of this is that in any problem where there is a
competitive algorithm, whose competitive factor does not increase with
problem size (such as most of the other examples where we have done
competitive analysis, such as maintaining a search list, or snoopy caching
problems), there will never be a necessity part of such a theorem.  For
these problems one could still envision the sufficiency part, and indeed,
an interesting problem is to consider this question for other on-line
problems, such as more general server problems.

By the way, I guess if you want to prove the ``necessity'' part of the
theorem for $r$-realizable sequences, then you'll have to let $k$ grow as
$e^m$, in order to produce a big enough problem so that the marking
algorithm doesn't undermine the proof.

Anyway, now you know my opinion.  What do you think?  I'll leave any further
editing of this section up to you.   D. Sleator
}

Algorithm $A$ will be a lazy algorithm; i.e., it will move a server only
when a vertex is requested that is currently not covered by a server. Let
$\sigma$ be a fixed request sequence. The time interval $(t_1,t_2)$ is
called a $v$-interval for $A$ if, when processing $\sigma$, $A$ moves a
server to vertex $v$ at time $t_1$, leaves the server on vertex $v$ until
time $t_2$, and then moves that server at time $t_2$. Algorithm $A$ is said
to {\it punish} algorithm $B$ at time $t_2$ if, for some $v$ and
$t_1$, $(t_1,t_2)$ is a
$v$-interval for $A$, and, for some $({t'}_1,{t'}_2)$ such that ${t'}_1 \leq
t_1 < {t'}_2 \leq t_2$, $({t'}_1,{t'}_2)$ is a $v$-interval
for $B$. Clearly, $C_B(\sigma)$ is at least as great as the number of time
steps at which $A$ punishes $B$. Thus, it suffices to show that $A$ can
punish each $B(i)$ at least $\lfloor C_A(\sigma)/c(i) \rfloor$ times.
We shall show that, at each step at which $A$ incurs a unit of cost, $A$ has
complete freedom to decide which algorithm $B(i)$ to punish. Let $S(A,t)$ be
the set of vertices that $A$ covers by servers just before request
$\sigma(t)$ arrives, and suppose that $\sigma(t) \not\in S(A,t)$, so that
$A$ must incur a unit of cost in order to process $\sigma(t)$. Let $S(B(i),
t+1)$ be the set of vertices that $B(i)$ covers by servers just after
processing $\sigma(t)$. Since the sets $S(A,t)$ and $S(B(i), t+1)$ each have
cardinality $k$, and since $S(B(i),t+1)$ contains $\sigma(t)$ but $S(A,t)$
does not, there must be some vertex $u$ that lies in $S(A,t)$ but does not
lie in $S(B(i), t+1)$. Then $A$ can punish $B(i)$ at step $t$ by moving a
server from vertex $u$ to vertex $\sigma(t)$.

Let $\mbox{PUN}(i,s,\sigma)$ denote the number of times $A$ punishes $B(i)$
while processing the first $s$ requests in $\sigma$,  and let
$C_A(s,\sigma)$ denote the cost that $A$ incurs during  the processing of
the first $s$ elements of $\sigma$. Suppose that $A$ must move a server in
order to process $\sigma(t)$. Then it chooses the server to move in such a
way as to punish that algorithm $B(i)$ for which $c(i)(\mbox{PUN}(i,t-1,
\sigma) +1)$ is least. It is easily verified that, provided $\sum{1/c(i)}
\leq 1$, the following holds for all positive integers $r$ and all $i$:
$B(i)$ gets punished at least $\lfloor r/c(i)\rfloor$ times by the time $A$
incurs a cost of $r$. This completes the proof of the sufficiency of
(\ref{X}).

(Necessity) Let $m$  be a positive  integer.  Let $c^* =
(c(1),c(2),...,c(m))$ be such that $\sum{1/c(i)} > 1$. We construct on-line
deterministic algorithms $B(1),B(2),...,B(m)$ of type $(2m-1,2m)$ such that
no on-line deterministic algorithm $A$ can be $c(i)$-competitive against
each $B(i)$.  Since $k = 2m-1$ and $n = 2m$, it will be the case that, at
any step in the execution of an on-line deterministic algorithm, exactly one
of the $2m$ possible vertices fails to be covered by a server.  For $i =
1,2,...,m$ let $B(i)$ be the algorithm that keeps all vertices except $i$
and $i+m$ permanently covered by servers, and thatshuttles the remaining server
between $i$ and $i+m$ in response to requests for those two vertices.
No two of these algorithms are ever required to move a server at the same
time. At any stage in the execution of a deterministic on-line algorithm $A$
there will exist some vertex that is not covered; this is true because there
are $2m$ vertices and only $2m-1$ servers. Thus, given any deterministic
on-line algorithm $A$ and any positive integer $N$, it is possible to
construct a request sequence $\tau(N)$ of length $N$ that causes $A$ to move
a server at every step. Then $C_A(\tau(N) = N$, and $\sum{C_{B(i)}(\tau(N))}
\leq N$. If $A$ is to be $c(i)$-competitive with each of of the on-line
algorithms $B(i)$ then there must exist constants $a(i)$ such that, for all
$i$ and all $N$,
$$
C_A(\sigma(N)) \leq c(i) \cdot C_{B(i)}(\sigma(N)) + a(i).
$$
But these inequalities,together with the fact that $\sum{1/c(i)} > 1$, lead to acontradiction for sufficiently large $N$.

\qed

We  now extend our definitions to the case of randomized algorithms.
Let $A$ and $B$ be randomized on-line algorithms of the same type. Let us
say that algorithm $A$ is {\it $c$-competitive} against algorithm $B$ if
there exists a constant $a$ such that, for every request sequence $\sigma$,
$$
\overline{C_A}(\sigma) \leq c\cdot \overline{C_B}(\sigma)  +  a.
$$
Let $c^*=(c(1),c(2),\ldots,c(m))$ be a sequence of positive reals. Then
$c^*$ is said to be {\it $r$-realizable} if for every type $(k,n)$, and for
every sequence $B(1),B(2),\ldots,B(m)$ of randomized on-line algorithms of
type $(k,n)$, there exists a randomized on-line algorithm $A$ of type
$(k,n)$ such that, for $i = 1,2,\ldots,m$, $A$ is $c(i)$-competitive against
$B(i)$.
\begin{theorem}
\label{realizabletheorem2}
If $c^*$ is realizable then $c^*$ is $r$-realizable.
\end{theorem}

\proof
Our proof is modeled after the proof of sufficiency in Theorem
\ref{realizabletheorem1}.  In that proof, deterministic on-line algorithms
$B(1),B(2),\ldots,B(m)$ of type $(k,n)$
were given, and the deterministic on-line algorithm
$A$ of type $(k,n)$
was constructed to be $c(i)$-competitive against $B(i)$ for each $i$.
The construction had the property that the action of $A$ in response to the
$t^{th}$ request in an input sequence $\sigma$ was completely determined by
$c^*$ and the actions of the algorithms $B(i)$ in response to the first $t$
requests in $\sigma$.  The construction ensures that, if $c^*$ is
realizable, then $B(i)$ incurs cost at least $\lfloor r/c(i) \rfloor$ by the
time $A$ incurs cost $r$, and hence $C_{B(i)}(\sigma) \geq \lfloor
C_A(\sigma)/c(i)\rfloor$. Let us call this construction PUNISH.

We shall extend PUNISH to the randomized case in a straightforward manner. A
randomized on-line algorithm may be viewed as basing its actions on the
request sequence $\sigma$ presented to it and on an infinite sequence $\rho$
of independent unbiased random bits. The action of the algorithm on
$\sigma(t)$, the $t^{th}$ request in $\sigma$, will be determined by the first
$t$ requests in $\sigma$ and by some initial part of the infinite sequence
$\rho$. Let $C_B(\sigma,\rho)$ be the cost incurred when algorithm $B$ is
executed on request sequence $\sigma$ using the sequence $\rho$ of random
bits. Then $\overline{C_B}(\sigma)$ is the expected value of
$C_B(\sigma,\rho)$.

Let $c^*$ be a realizable sequence, and let $B(1),B(2),\ldots,B(m)$ be
randomized on-line algorithms. We shall construct a randomized on-line
algorithm $A$ of the same type that is $c(i)$-competitive against
$B(i)$, for $i = 1,2,\ldots,m$. We begin by giving a conceptual view
of Algorithm $A$, ignoring questions of effectiveness.  Algorithm $A$
starts by constructing an infinite sequence $\rho$ of independent,
unbiased bits. Then, as successive requests in the input sequence
$\sigma$ arrive, it calculates the actions of each of the $B(i)$ on
these requests in $\sigma$ using the sequence of random bits $\rho$,
and determines its own actions by applying PUNISH to $c^*$ and the
actions of $B(1),B(2),\ldots,B(m)$ on $\sigma$ with random bits
$\rho$. This ensured that, for $i=1,2,\ldots,m$,
$C_{B(i)}(\sigma,\rho) \geq \lfloor C_A(\sigma,\rho)/c(i)\rfloor$,
and, averaging over all choices of the random bits $\rho$,
$\overline{C_A}(\sigma) \leq c(i)\cdot\overline{C_B}(\sigma) + a(i)$.

\qed

To do this simulation it is not actually necessary for $A$ to generate
an infinite sequence $\rho$.  To process requests $1,2,\ldots,t$, $A$
needs to generate as many random bits as are required by any $B(i)$.
Algorithm $A$ must also remember that portion of the sequence $\rho$
that it has given to some $B(i)$, but not all of them.  Of course $A$
is also required to simulate the behavior of each $B(i)$ on
the given input sequence.

\ignore{
If the algorithms $B(i)$ are presented in a computationally effective
manner then $A$ will need to generate only a finite number of random
bits (i.e., a finite initial part of the infinite sequence $\rho$) to
determine its own behavior on $\sigma$.}
\ignore{; in this sense theconstruction of $A$ is effective.}

\section{Extensions}

The problem of devising a strongly competitive algorithm for any $k$
and $n$ was solved by McGeoch and Sleator \cite{ms}.  Their
{\it partitioning algorithm\/} is much more
complicated than the marking algorithm, but achieves the optimal
competitive factor of $H_k$.

For deterministic server problems all evidence indicates that the
optimal competitive factor is $k$, and is therefore independent of the
distances in the graph \cite{ckpv,cl,mmspaper}.  This is not true in
the randomized case.  Karlin {\it et al.} \cite{kmmo} have shown that
for two servers in a graph that is an isosceles triangle the best
competitive factor that can be achieved is a constant that approaches
$e/(e-1) \cong 1.582$ as the length of the similar sides go to
infinity.  This contrasts with the uniform 3-vertex, 2-server problemfor which the
marking algorithm is $1.5$-competitive.  Analyzing the competitiveness
of other non-uniform problems remains a challenging open problem.

\ignore{
The most outstanding open problem involving randomized algorithms for
server problems is to generalize our results for the uniform case to
to non-uniform problems.  The only non-uniform case that has been
analyzed exactly is two servers on an isosceles triangle \cite{kmmo}.
Karlin, Manasse, McGeoch, and Owicki showed that as the length of the
two similar sides of the triangle approach infinity, the best
competitive factor that can be achieved approaches $e/(e-1)$.}

Sleator and Tarjan \cite{st} used a slightly different framework to study
competitiveness in paging problems.  They compared on-line
algorithms to off-line algorithms with different numbers of servers
(amounts of fast memory).  They showed that LRU running with $k$
servers performs within a factor of $k/(k-h+1)$ of any off-line
algorithm with $h \leq k$ servers, and that this is the minimum
competitive factor that can be achieved.  Young \cite{young} has
extended this analysis to randomized algorithms.  He has shown that
the marking algorithm is roughly $2 \ln (k /(k-h+1))$-competitiveunder these circumstances.
There are many open problems involving the combining of on-line algorithms.
Most notable of these is to extend the technique of constructing an
algorithm competitive with several others to other problems besides the
uniform server problem.  Candidates include non-uniform server problems,
maintaining a list \cite{st}, and snoopy caching \cite{kmrs}.

\ignore{open question about combining algorithms is to fully characterize the
$r$-realizable sequences.}
\ignore{
We have left the obvious open problem of precisely characterizing the
vectors of constants $c(1),\ldots,c(m)$ such that
}

\section*{Acknowledgement}
The authors would like to thank Jorge Stolfi and an anonymous referee
for many helpful suggestions.

\bibliographystyle{plain}

\end{document}